# New Comparative Study Between DES, 3DES and AES within Nine Factors

Hamdan.O.Alanazi, B.B.Zaidan, A.A.Zaidan, Hamid A.Jalab, M.Shabbir and Y. Al-Nabhani

ABSTRACT---With the rapid development of various multimedia technologies, more and more multimedia data are generated and transmitted in the medical, also the internet allows for wide distribution of digital media data. It becomes much easier to edit, modify and duplicate digital information .Besides that, digital documents are also easy to copy and distribute, therefore it will be faced by many threats. It is a big security and privacy issue, it become necessary to find appropriate protection because of the significance, accuracy and sensitivity of the information. , which may include some sensitive information which should not be accessed by or can only be partially exposed to the general users. Therefore, security and privacy has become an important. Another problem with digital document and video is that undetectable modifications can be made with very simple and widely available equipment, which put the digital material for evidential purposes under question. Cryptography considers one of the techniques which used to protect the important information. In this paper a three algorithm of multimedia encryption schemes have been proposed in the literature and description. The New Comparative Study between DES, 3DES and AES within Nine Factors achieving an efficiency, flexibility and security, which is a challenge of researchers.

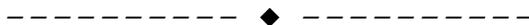

## 1. INTRODUCTION

Cryptography is the art and science of protecting information from undesirable individuals by converting it into a form non-recognizable by its attackers while stored and transmitted [1]. Data cryptography mainly is the scrambling of the content of data, such as text, image, audio, video and so forth to make the data unreadable, invisible or unintelligible during transmission or storage called Encryption. The main goal of cryptography is keeping data secure form unauthorized attackers. The reverse of data encryption is data Decryption,

- *Hamdan.O.Alanazi – Master Student, Department of Computer System & Technology, University Malaya, Kuala Lumpur, Malaysia.*
- *B. B. Zaidan – PhD Candidate on the Department of Electrical & Computer Engineering / Faculty of Engineering, Multimedia University, Cyberjaya, Malaysia.*
- *A. A. Zaidan – PhD Candidate on the Department of Electrical & Computer Engineering , Faculty of Engineering , Multimedia University , Cyberjaya, Malaysia.*
- *Dr. Hamid.A.Jalab- Senior Lecturer, Department of Computer Science &Information Technology, University Malaya, Kuala Lumpur, Malaysia*
- *M.Shabbir – Master Student, Department of Computer System & Technology, University Malaya, Kuala Lumpur, Malaysia*
- *Y. Al-Nabhani – Master Student, Department of Computer System & Technology, University Malaya, Kuala Lumpur, Malaysia.*

In modern days cryptography is no longer limited to secure sensitive military information but recognized as one of the major components of the security policy of any organization and considered industry standard for providing information security, trust, controlling access to resources, and electronic financial transactions. The which recuperate the original data. Since cryptography first known usage in ancient Egypt it has passed through different stages and was affected by any major event that affected the way people handled information. In the World War II for instance cryptography played an important role and was a key element that gave the allied forces the upper hand, and enables them to win the war sooner, when they were able to dissolve the Enigma cipher machine which the Germans used to encrypt their military secret communications [2].

Original data that to be transmitted or stored is called plaintext, the one that can be readable and understandable either by a person or by a computer. Whereas the disguised data so-called ciphertext, which is unreadable, neither human nor machine can properly process it until it is decrypted. A system or product that provides encryption and decryption is called cryptosystem [3]. Cryptosystem uses an encryption algorithms which determines how simple or complex the encryption process will be, the necessary software component, and the key (usually a long string of bits), which works with the algorithm to encrypt and decrypt the data [3], [4]. In the 19th century, a famous theory about the security principle of any encryption system has been proposed by Kerchhoff. This theory has become the most important principle in designing a cryptosystem for researchers and engineers. Kirchhoff observed that the encryption algorithms are supposed to be known to the opponents [5]. Thus, the security of an encryption system should rely on the secrecy of the encryption/decryption key instead of the encryption algorithm itself. For even though in the very beginning the opponent doesn't know the algorithm, the encryption system will not be able to protect the ciphertext once the algorithm is broken. The security level of an encryption algorithm is measured by the size of its key space [6]. The larger size of the key space is, the more time the attacker needs to do the exhaustive search of the key space, and thus the higher the security level is. In encryption, the key is piece of information (value of comprise a large sequence of random bits) which specifies the particular transformation of plaintext to ciphertext, or vice versa during decryption. Encryption key based on the key space, which is the range of the values that can be used to assemble a key. The larger key space the more possible keys can be constructed (e.g. today we commonly use key sizes of 128,192,or 256 bit , so the key size of 256 would provide a $2^{256}$ key space) [5],[6]. The strength of the encryption algorithm relies on the secrecy of the key, length of the key, the initialization vector, and how they all work together [6]. Depend on the algorithm, and length of the key, the strength of encryption can be considered. Assume that if the key can be broken in three hours using Pentium 4 processor the cipher consider is not strong at all, but if the key can broken with thousand of multiprocessing systems within a million years, then the cipher is pretty darn strong. There are two encryption/decryption key types: In some of encryption technologies when two end points need to communicate with one another via encryption, they must use the same algorithm, and in the most of the time the same key, and in other encryption technologies, they must use different but related keys for encryption and decryption purposes. Cryptography algorithms are either symmetric algorithms, which use symmetric keys (also called secret keys), or asymmetric algorithms, which use asymmetric keys (also called public and private keys) [7].

## 2. CRYPTOGRAPHY WITH BLOCK CIPHER

In Cryptography, a block cipher is a symmetric key cipher which operates on fixed-length groups of bits, termed blocks, with an unvarying transformation. When encrypting, a block cipher might take a (for example) 128-bit block of plaintext as input, and outputs a corresponding 128-bit block of cipher text. The exact transformation is controlled using a second input — the secret key [7]. Decryption is similar: the decryption algorithm takes, in this example, a 128-bit block of cipher text together with the secret key, and yields the original 128-bit block of plaintext. To encrypt messages longer than the block size (128 bits in the above example), a mode of operation is used. Block ciphers can be contrasted with stream ciphers; a stream cipher operates on individual digits one at a time and the transformation varies during the encryption. The distinction between the two types is not always clear-cut: a block cipher, when used in certain modes of operation, acts effectively as a stream cipher as shown in Fig 1.

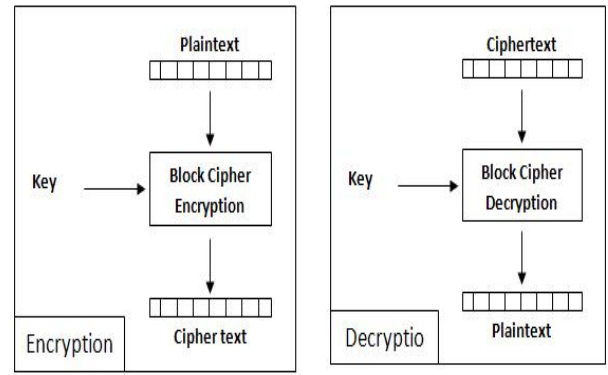

Fig 1. Stream Cipher.

An early and highly influential block cipher design is the Data Encryption Standard (DES). The (DES) is a cipher (a method for encrypting information) Selected as an official Federal Information Processing Standard (FIPS) for the United States in 1976, and which has subsequently enjoyed widespread use internationally. The algorithm was initially controversial, with classified design elements, a relatively short key length, and suspicions about a National Security Agency (NSA) backdoor. DES consequently came under intense academic scrutiny, and motivated the modern understanding of block ciphers and their cryptanalysis. DES is now considered to be insecure for many applications. This is chiefly due to the 56-bit key size being too small; DES keys have been broken in less than 24 hours. There are also some analytical results which demonstrate theoretical weaknesses in the cipher, although they are infeasible to mount in practice. The algorithm is believed to be practically secure in the form of Triple DES, although there are theoretical attacks. In recent years, the cipher has been superseded by the Advanced Encryption Standard [8].

### 2.1 Data Encryption Standard (DES)

DES is a Feistel-type Substitution-Permutation Network (SPN) cipher, specified in FIPS PUB 46. The result of a 1970s effort to produce a U.S. encryption standard. DES uses a 56-bit key which can be broken using brute-force methods, and is now considered obsolete. A 16 cycle Feistel system is used, with an overall 56-bit key permuted into 16 48-bit subkeys, one for each cycle. To decrypt, the identical algorithm is used, but the order of subkeys is reversed. The L and R blocks are 32 bits each, yielding an overall block size of 64 bits. The hash function "f", specified by the standard using the so-called "S-boxes", takes a 32-bit data block and one of the 48-bit subkeys as input and produces 32 bits of output. Sometimes DES is said to use a 64-bit key, but 8 of the 64 bits are used only for parity checking, so the effective key size is 56 bits [9].

Since the time DES was adopted (1977), it has been widely speculated that some kind of backdoor was designed into the cryptic S-boxes, allowing those "in the know" to effectively crack DES. Time has proven such speculation idle. Regardless of any backdoors in the hash function, the rapid advances in the speed of electronic circuitry over the last 20 years, combined with the natural parallelism of Feistel ciphers and DES's relatively small key size, have rendered the algorithm obsolete. In 1998, the Electronic Frontier Foundation built a DES Cracker (full specifications available online) for less than $250,000 that can decode DES messages in less than a week [7],[8],[9].

**2.2 Triple DES**

Triple DES was developed to address the obvious flaws in DES without designing a whole new cryptosystem. Triple DES simply extends the key size of DES by applying the algorithm three times in succession with three different keys. The combined key size is thus 168 bits (3 times 56), beyond the reach of brute-force techniques such as those used by the EFF DES Cracker. Triple DES has always been regarded with some suspicion, since the original algorithm was never designed to be used in this way, but no serious flaws have been uncovered in its design, and it is today available cryptosystem used in a number of Internet protocols[8],[9].

**2.3 Advanced Encryption Standard (AES) / Rijndael**

In the late 1990s, the U.S. National Institute of Standards and Technology (NIST) conducted a competition to develop a replacement for DES. The winner, announced in 2001, is the Rijndael (pronounced "rhine-doll") algorithm, destined to become the new Advanced Encryption Standard. Rijndael mixes up the SPN model by including Galios field operations in each round. Somewhat similar to RSA modulo arithmetic operations, the Galios field operations produce apparent gibberish, but can be mathematically inverted.AES have Security is not an absolute; it's a relation between time and cost. Any question about the security of encryption should be posed in terms of how long time, and how high cost will it take an attacker to find a key?

Currently, there are speculations that military intelligence services possibly have the technical and economic means to attack keys equivalent to about 90 bits, although no civilian researcher has actually seen or reported of such a capability. Actual and demonstrated systems today, within the bounds of a commercial budget of about 1 million dollars can handle key lengths of about 70 bits. An aggressive estimate on the rate of technological progress is to assume that technologies will double the speed of computing devices every year at an unchanged cost. If correct, 128-bit keys would be in theory be in range of a military budget within 30-40 years. An illustration of the current status for AES is given by the following example, where we assume an attacker with the capability to build or purchase a system that tries keys at the rate of one billion keys per second. This is at least 1 000 times faster than the fastest personal computer in 2004. Under this assumption, the attacker will need about 10 000 000 000 000 000 000 years to try all possible keys for the weakest version, AES-128.The key length should thus be chosen after deciding for how long security is required, and what the cost must be to brute force a secret key. In some military circumstances a few hours or days security is sufficient – after that the war or the mission is completed and the information uninteresting and without value. In other cases a lifetime may not be long enough. There is currently no evidence that AES has any weaknesses making any attack other than exhaustive search, i.e. brute force, possible. Even AES-128 offers a sufficiently large number of possible keys, making an exhaustive search impractical for many decades, provided no technological breakthrough causes the computational power available to increase dramatically and that theoretical research does not find a short cut to bypass the need for exhaustive search. There are many pitfalls to avoid when encryption is implemented, and keys are generated. It is necessary to ensure each and every implementations security, but hard since it requires careful examination by experts. An important aspect of an evaluation of any specific implementation is to determine that such an examination has been made, or can be conducted [10],[11].

**3. COMPARISON BETWEEN AES, 3DES AND DES.**

Advance Encryption Standard (AES) and Triple DES (TDES or 3DES) are commonly used block ciphers. Whether you choose AES or 3DES depend on your needs. In this section it would like to highlight their differences in terms of security and performance (Seleborg, 2004).Since 3DES is based on DES algorithm, it will talk about DES first. DES was developed in 1977 and it was carefully designed to work better in hardware than software. DES performs lots of bit manipulation in substitution and permutation boxes in each of 16 rounds. For example, switching bit 30 with 16 is much simpler in hardware than software. DES encrypts data in 64 bit block size and uses effectively a 56 bit key. 56 bit key space amounts to approximately 72 quadrillion possibilities. Even though it seems large but according to today's computing power it is not sufficient and vulnerable to brute force attack. Therefore, DES could not keep up with advancement in technology and it is no longer appropriate for security. Because DES was widely used at that time, the quick solution was to introduce 3DES which is secure enough for most purposes today.3DES is a construction of applying DES three times in sequence. 3DES with three different keys (K1, K2 and K3) has effective key length is 168 bits (The use of three distinct key is recommended of 3DES.). Another variation is called two-key (K1 and K3 is same) 3DES reduces the effective key size to 112 bits which is less secure. Two-key 3DES is widely used in electronic payments industry. 3DES takes three times as much CPU power than compare with its predecessor which is significant performance hit. AES outperforms 3DES both in software and in hardware [12],[13].The

Rijndael algorithm has been selected as the Advance Encryption Standard (AES) to replace 3DES. AES is modified version of Rijndael algorithm.Advance Encryption Standard evaluation criteria among others was [8],[9],[10],[11]:
• Security
• Software & Hardware performance
• Suitability in restricted-space environments
• Resistance to power analysis and other implementation attacks.

Rijndael was submitted by Joan Daemen and Vincent Rijmen.When considered together Rijndael's combination of security, performance, efficiency, implementability, and flexibility made it an appropriate selection for the AES.By design AES is faster in software and works efficiently in hardware. It works fast even on small devices such as smart phones; smart cards etc.AES provides more security due to larger block size and longer keys.AES uses 128 bit fixed block size and works with 128, 192 and 256 bit keys.Rigndael algorithm in general is flexible enough to work with key and block size of any multiple of 32 bit with minimum of128 bits and maximum of 256 bits.AES is replacement for 3DES according to NIST both ciphers will coexist until the year2030 allowing for gradual transition to AES.Even though AES has theoretical advantage over 3DES for speed and efficiency in some hardware implementation 3DES may be faster where support for 3DES is mature

Table 1
Comparison between AES , 3DES and 3DES.

| Factors | AES | 3DES | DES |
|---|---|---|---|
| Key Length | 128, 192, or 256 bits | (k1,k2 and k3) 168 bits (k1 and k2 is same) 112bits | 56 bits |
| Cipher Type | Symmetric block cipher | Symmetric block cipher | Symmetric block cipher |
| Block Size | 128, 192, or 256 bits | 64bits | 64 bits |
| Developed | 2000 | 1978 | 1977 |
| Cryptanalysis resistance | Strong against differential, truncated differential, linear, interpolation and square attacks | Vulnerable to differential, Brute Force attacker could be analyze plaint text using differential cryptanalysis. | Vulnerable to differential and linear cryptanalysis; weak substitution tables |
| Security | Considered secure | one only weak which is Exit in DES. | Proven inadequate |
| Possible Keys | $2^{128}$, $2^{192}$, or $2^{256}$ | $2^{112}$ or $2^{168}$ | $2^{56}$ |
| Possible ASCII printable character keys | $95^{16}$, $95^{24}$, or $95^{32}$ | $95^{14}$ or $95^{21}$ | $95^{7}$ |
| Time required to check all possible keys at 50 billion keys per second** | For a 128-bit key: $5 \times 10^{21}$ years | For a 112-bit key: 800 Days | For a 56-bit key: 400 Days |

## 4. CONCLUSION

In this paper a new comparative study between DES, 3DES and AES were presented in to nine factors, Which are key length, cipher type, block size, developed, cryptanalysis resistance, security, possibility key, possible ACSII printable character keys, time required to check all possible key at 50 billion second, these eligible's proved the AES is better than DES and 3DES.

## ACKNOWLEDGEMENT

Thanks in advance for the entire worker in this project, and the people who support in any way, also I want to thank University of Malay for the support they offered, also I would like to extend our deep apparitions and thanks Dr.Iman for his support.

**Hamdan Al-Anazi**: has obtained his bachelor dgree from "King Suad University", Riyadh, Saudi Arabia. He worked as a lecturer at Health College in the Ministry of Health in Saudi Arabia, then he worked as a lecturer at King Saud University in the computer department. Currently he is Master candidate at faculty of Computer Science & Information Technology at University of Malaya in Kuala Lumpur, Malaysia. His research interest on Information Security, cryptography, steganography and digital watermarking, He has contributed to many papers some of them still under reviewer.

**Bilal Bahaa Zaidan:** He obtained his bachelor degree in Mathematics and Computer Application from Saddam University/Baghdad followed by master in data communication and computer network from University of Malaya. He led or member for many funded research projects and He has published more than 55 papers at various international and national conferences and journals, His interest area are Information security (Steganography and Digital watermarking), Network Security (Encryption Methods) , Image Processing (Skin Detector), Pattern Recognition , Machine Learning (Neural Network, Fuzzy Logic and Bayesian) Methods and Text Mining and Video Mining. .Currently, he is PhD Candidate on the Department of Electrical & Computer Engineering / Faculty of Engineering / Multimedia University / Cyberjaya, Malaysia. He is members IAENG, CSTA, WASET, and IACSIT. He is reviewer in the (IJSIS, IJCSN, IJCSE and IJCIIS).

**Aos Alaa Zaidan**: He obtained his 1st Class Bachelor degree in Computer Engineering from university of Technology / Baghdad followed by master in data communication and computer network from University of Malaya. He led or member for many funded research projects and He has published more than 55 papers at various international and national conferences and journals, His interest area are Information security (Steganography and Digital watermarking), Network Security (Encryption Methods) , Image Processing (Skin Detector), Pattern Recognition , Machine Learning (Neural Network, Fuzzy Logic and Bayesian) Methods and Text Mining and Video Mining. .Currently, he is PhD Candidate on the Department of Electrical & Computer Engineering / Faculty of Engineering / Multimedia University / Cyberjaya, Malaysia. He is members IAENG, CSTA, WASET, and IACSIT. He is reviewer in the (IJSIS, IJCSN, IJCSE and IJCIIS).

**Dr.Hamid.A.Jalab**: Received his B.Sc degree from University of Technology, Baghdad, Iraq. MSc & Ph.D degrees from Odessa Polytechnic National State University 1987 and 1991, respectively. Presently, Visiting Senior Lecturer of Computer System and Technology, Faculty of Computer Science and Information Technology, University of Malaya, Malaysia. various international and national conferences and journals, His interest area are Information security (Steganography and Digital watermarking), Network Security (Encryption Methods) , Image Processing (Skin Detector), Pattern Recognition , Machine Learning (Neural Network, Fuzzy Logic and Bayesian) Methods and Text Mining and Video Mining.

**Mohamed Shabbir -** he is master student in the Faculty of Computer Science and Information Technology/University of Malaya /Kuala Lumpur/Malaysia, He has contribution for many papers at international conferences and journals.

**Yahya Al-Nabhani -** he is master student in the Faculty of Computer Science and Information Technology/University of Malaya /Kuala Lumpur/Malaysia, He has contribution for many papers at international conferences and journals.